\documentclass[journal]{IEEEtran}
\usepackage[table,xcdraw]{xcolor}
\usepackage{amsmath,amssymb,amsfonts}
\usepackage{algorithmic}
\usepackage{graphicx}
\usepackage{textcomp}

\usepackage{etoolbox}
\makeatletter
\@ifundefined{color@begingroup}%
{\let\color@begingroup\relax
\let\color@endgroup\relax}{}%
\def\fix@ieeecolor@hbox#1{%
\hbox{\color@begingroup#1\color@endgroup}}
\patchcmd\@makecaption{\hbox}{\fix@ieeecolor@hbox}{}{\FAILED}
\patchcmd\@makecaption{\hbox}{\fix@ieeecolor@hbox}{}{\FAILED}
\definecolor{color}{rgb}{0, 0, 0}
\definecolor{color_revise}{rgb}{0, 0, 0}

\usepackage{soul}
\definecolor{colDim1}{HTML}{EAF1FB}
\definecolor{colTrad1}{HTML}{FDF2E9}
\definecolor{colLAM1}{HTML}{E7F4E4}
\definecolor{headerBlue}{HTML}{B0D0FF}

\definecolor{colDim2}{HTML}{EDF7EB}
\definecolor{colTrad2}{HTML}{EDF7EB}
\definecolor{colLAM2}{HTML}{EAF5E4}
\definecolor{headerGreen}{HTML}{B7DAB1}

\usepackage{soul}
\usepackage{array}
\usepackage[ruled, linesnumbered, vlined]{algorithm2e}
\usepackage{textcomp}
\usepackage{multirow}
\usepackage{bm}
\usepackage{booktabs}
\usepackage{ntheorem}
\usepackage{subfigure} 
\usepackage{orcidlink} 
\hypersetup{hidelinks}
\usepackage{mathrsfs}
\usepackage{mathtools}
\usepackage{epstopdf}
\usepackage{cases}
\def\BibTeX{{\rm B\kern-.05em{\sc i\kern-.025em b}\kern-.08em
		T\kern-.1667em\lower.7ex\hbox{E}\kern-.125emX}}
\usepackage{cases}
\theoremseparator{.}

\theorembodyfont{}

\begin{document}

\title{Large AI Model-Enabled Secure Communications in Low-Altitude Wireless Networks: Concepts, Perspectives and Case Study}

\author{Chuang~Zhang$^{\orcidlink{0000-0002-0505-0512}}$,\IEEEmembership{}
        Geng~Sun$^{\orcidlink{0000-0001-7802-4908}}$,~\IEEEmembership{Senior Member,~IEEE,}
        Yijing~Lin$^{\orcidlink{0000-0003-2702-7679}}$,~\IEEEmembership{Member,~IEEE,}\\
        Weijie~Yuan$^{\orcidlink{0000-0002-2158-0046}}$,~\IEEEmembership{Senior Member,~IEEE,}
        Sinem~Coleri$^{\orcidlink{0000-0002-7502-3122}}$, \IEEEmembership{Fellow,~IEEE,}
        and Dusit~Niyato$^{\orcidlink{0000-0002-7442-7416}}$,~\IEEEmembership{Fellow,~IEEE}
        \thanks{This study is supported in part by the National Natural Science Foundation of China (62272194, 62471200), in part by the Science and Technology Development Plan Project of Jilin Province (20250101027JJ), and in part by Seatrium New Energy Laboratory, Singapore Ministry of Education (MOE) Tier 1 (RT5/23 and RG24/24), the Nanyang Technological University (NTU) Centre for Computational Technologies in Finance (NTU-CCTF), and the Research Innovation and Enterprise (RIE) 2025 Industry Alignment Fund - Industry Collaboration Projects (IAF-ICP) (Award I2301E0026), administered by Agency for Science, Technology and Research (A*STAR). \textit{(Corresponding author: Geng Sun.)}}
	\IEEEcompsocitemizethanks{
            \IEEEcompsocthanksitem Chuang Zhang is with the College of Computer Science and Technology, Jilin University, Changchun 130012, China, and also with the Singapore University of Technology and Design, Singapore 487372 (e-mail: chuangzhang1999@gmail.com).
            \IEEEcompsocthanksitem Geng Sun is with the College of Computer Science and Technology, Key Laboratory of Symbolic Computation and Knowledge Engineering of Ministry of Education, Jilin University, Changchun 130012, China, and also with the College of Computing and Data Science, Nanyang Technological University, Singapore 639798 (e-mail: sungeng@jlu.edu.cn).
            \IEEEcompsocthanksitem Yijing Lin is with the State Key Laboratory of Networking and Switching Technology, Beijing University of Posts and Telecommunications, Beijing 100876, China. (e-mail: yjlin@bupt.edu.cn)
            \IEEEcompsocthanksitem  Weijie Yuan is with the School of Automation and Intelligent Manufacturing, Southern University of Science and Technology, Shenzhen 518055, China (e-mail: yuanwj@sustech.edu.cn).
            \IEEEcompsocthanksitem Sinem Coleri is with the Department of Electrical and Electronics Engineering, Koc University, 34450 Istanbul, Turkey (e-mail: scoleri@ku.edu.tr).
            \IEEEcompsocthanksitem Dusit Niyato is with the College of Computing and Data Science, Nanyang Technological University, Singapore 639798 (e-mail: dniyato@ntu.edu.sg). 
}

}

\markboth{Journal of \LaTeX\ Class Files,~Vol.~X, No.~X, July~2025}%
{Shell \MakeLowercase{\textit{et al.}}: Bare Demo of IEEEtran.cls for Computer Society Journals}

\IEEEtitleabstractindextext{%
	\begin{abstract}		
		Low-altitude wireless networks (LAWNs) have the potential to revolutionize communications by supporting a range of applications, including urban parcel delivery, aerial inspections and air taxis. However, compared with traditional wireless networks, LAWNs face unique security challenges due to low-altitude operations, frequent mobility and reliance on unlicensed spectrum, making it more vulnerable to some malicious attacks. In this paper, we investigate some large artificial intelligence model (LAM)-enabled solutions for secure communications in LAWNs. Specifically, we first explore the amplified security risks and important limitations of traditional AI methods in LAWNs. Then, we introduce the basic concepts of LAMs and delve into the role of LAMs in addressing these challenges. {\color{color_revise}{To demonstrate the practical benefits of LAMs for secure communications in LAWNs, we propose a novel LAM-based optimization framework. This framework uses chain-of-thought-enabled large language models (LLMs) to enhance state features derived from handcrafted representations and design intrinsic rewards based on these enhanced features. This approach improves reinforcement learning performance for secure communication tasks.}} Through a typical case study, simulation results validate the effectiveness of the proposed framework. Finally, we outline future directions for integrating LAMs into secure LAWN applications.
	\end{abstract}
	
	}

\maketitle
\IEEEdisplaynontitleabstractindextext
\IEEEpeerreviewmaketitle
\begin{figure*}
    \centering
    \includegraphics[width=0.95\linewidth]{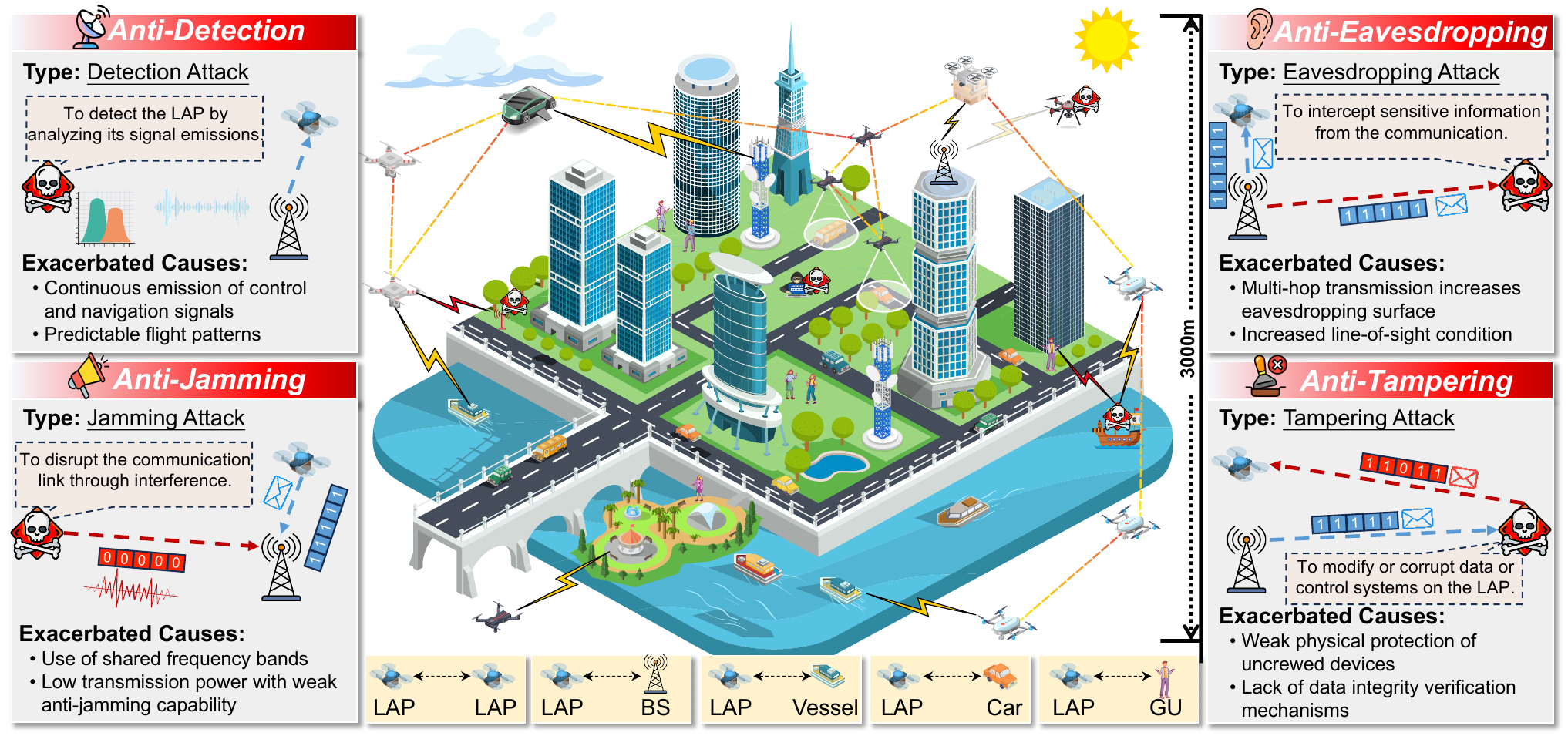}
    \caption{{\color{color_revise}{A systematic overview of communication security threats to LAPs in LAWNs.}}}
    \label{Fig: Mag_Fig_1}
    \vspace{-14pt}
\end{figure*}
%
%
\vspace{-12pt}
\section{Introduction}
\label{Sec: Introduction}

\par As low-altitude economic applications continue to advance, low-altitude wireless networks (LAWNs) are emerging to become a crucial component of next-generation wireless networks, providing essential support for applications such as urban parcel delivery, aerial inspections, and air taxis {\color{color_revise}{\cite{Wang2025}}}. Specifically, LAWNs offer unique advantages by utilizing high mobility, rapid deployment and flexible coverage of low-altitude platforms (LAPs), such as uncrewed aerial vehicles (UAVs) and electric vertical take-off and landing aircraft (eVTOLs), thereby allowing for seamless communication in dynamic low-altitude aerial environments.

\par However, the distinct characteristics of LAWNs give rise to communication security challenges that differ from traditional terrestrial wireless networks {\color{color_revise}{\cite{Cai2025}}}. {\color{color_revise}{Specifically, the dependence on LAPs introduces heightened vulnerability to eavesdropping and detection attacks due to increasing likelihood of line-of-sight exposure. Moreover, the inherent mobility and dynamic operational environments of LAPs pose significant challenges to network management, thereby complicating the maintenance of data integrity. Furthermore, the use of unlicensed spectrum exacerbates the issue by increasing susceptibility to interference from unauthorized entities.}} Therefore, these factors significantly amplify the risks of detection, eavesdropping, jamming and tampering in LAWNs.

\par To address these communication security challenges, traditional AI methods have been applied in various forms, which include supervised learning, unsupervised learning, and reinforcement learning (RL). These techniques have shown potential in detecting and mitigating security threats, such as identifying jamming signals, decreasing eavesdropping channel capacity and detecting anomalous behaviors. However, traditional AI methods have several limitations when it comes to dynamic and complex environments such as LAWNs. First, many of these methods have been designed for specific and narrow tasks, thus struggling to generalize across diverse scenarios. Moreover, traditional AI models typically rely on single-modality inputs, failing to leverage the multi-dimensional data that is crucial in LAWN environments, such as communication signals, environmental context and platform behaviors. Furthermore, these methods are reactive to attacks after they have been launched rather than predicting and mitigating potential threats proactively.

\par Fortunately, large artificial intelligence models (LAMs) offer a promising solution to these limitations \cite{Jiang2025}. {\color{color_revise}{Specifically, LAMs offer stronger adaptability, multi-modal intelligence integration, and long-term scalability, thereby enabling them to address the security demands of LAWNs more effectively than traditional AI approaches. For example, the WirelessGPT framework \cite{Yang2025} shows that LAM-based wireless models can perform effectively even under challenging conditions, such as a –5 dB SNR. In channel estimation, WirelessGPT achieves a 30\% reduction in normalized mean squared error compared to traditional methods, clearly demonstrating the performance improvements enabled by LAMs.}} Therefore, this paper aims to investigate the integration of LAMs into secure communications for LAWNs and our contributions are listed as follows.

\begin{itemize}
    \item We present a comprehensive overview of the unique security challenges in LAWNs and identify the fundamental limitations inherent in traditional AI-based methods.
    \item We examine the integration of LAMs into secure communications in LAWNs from both architectural and functional perspectives, highlighting their ability to enhance secure communications in LAWNs.
    \item {\color{color_revise}{We present an chain-of-thought-(CoT)-LAM-enhanced optimization framework for improving state representation and designing intrinsic rewards in RL toward secure communications in LAWNs. Simulation results demonstrate the superior performance of the proposed framework.}}
\end{itemize}

\vspace{-5pt}
%
%
\section{Secure Communications in LAWNs}
\label{Sec: Secure Communications in LAWNs}


\subsection{Background and Communication Security Risks of LAWNs}
\label{SubSec: Background of LAWNs}

\par LAWNs refer to a class of wireless communication systems that operate in the low-altitude airspace within the range of 0 to 3000 meters above sea level. As shown in Fig.~\ref{Fig: Mag_Fig_1}, {\color{color_revise}{these security risks are more severe than in terrestrial networks, which can be listed as follows.

\begin{itemize}
    \item \textbf{\textit{Malicious communication detection}:} Detection attacks aim to locate and track transmitters by analyzing their emitted signals. Under the requirement for flight stabilization and coordination, the frequent transmission of control and navigation data makes such emissions persistent and distinguishable in LAWNs, making them highly susceptible to detection attacks. Moreover, low-altitude operations and predictable trajectories introduce spatio-temporal patterns that adversaries can exploit to infer the presence of LAPs.

    \item \textbf{\textit{Transmission eavesdropping}:} Eavesdropping attacks seek to intercept sensitive information from communications. In LAWNs, low-altitude operations often increase the likelihood of line-of-sight channels \cite{Zhang2025}. Moreover, the use of multi-hop relay architectures also introduces additional communication links, which together increase the exposure surface of the network. As a result, adversaries have more opportunities to intercept transmitted signals at different points along the path, thereby increasing the overall risk of eavesdropping.

    \item \textbf{\textit{Transmission disruptions}:} Jamming attacks aim to disrupt communications by injecting interference into the transmission medium. In LAWNs, the lack of unified spectrum allocation and authentication standards, together with their relatively low transmit power due to energy constraints, makes channels vulnerable to interference. Consequently, adversaries can effectively launch jamming attacks that degrade legitimate channel capacity, especially in congested environments with limited spectrum.

    \item \textbf{\textit{Data tampering}:} Tampering attacks involve modifying, injecting or corrupting transmitted data to compromise the integrity or functionality of the system. In LAWNs, this threat is exacerbated by the limited physical protection of LAP devices and the lack of robust information integrity verification mechanisms. As such, adversaries may exploit these weaknesses to alter control or telemetry messages, thereby threatening flight safety and mission reliability.
\end{itemize}}}
\vspace{-12pt}
\subsection{Traditional AI-enabled Solutions}
\label{Subsec: Traditional AI-enabled Solutions for Secure Communications in LAWNs}

\par In response to the increasing complexity of security threats in LAWNs, traditional AI techniques have been actively explored as practical tools. Based on their learning paradigms, existing AI-enabled solutions can be categorized as follows. 

\begin{itemize}
    \item \textbf{\textit{Supervised learning for secure communications}:} One widely adopted approach in traditional AI is supervised learning, where models are trained on labeled data to perform classification or regression tasks. For example, the authors in \cite{Vo2025} proposed a UAV equipped with a reconfigurable intelligent surface-enabled cognitive radio system, where a supervised learning-based deep neural network model was employed to accomplish the throughput maximization of the covert secondary receivers.
    \item \textbf{\textit{Unsupervised learning for secure communications}:} In contrast, unsupervised learning offers the advantage of discovering hidden patterns or representations from unlabeled data, which is especially valuable when labeled datasets are scarce. Aiming to maximize the minimum achievable secrecy rate in UAV-assisted secure wireless-powered communication systems, Heo \textit{et al}. \cite{Heo2024} proposed an unsupervised deep neural network model with a tailored loss function regarding the secure channel capacity to efficiently approximate optimal schedule, power and trajectory control strategies of the UAV.
    \item \textbf{\textit{RL for secure communications}:} In addition to passive learning paradigms, RL offers an interactive framework where agents learn optimal strategies by interacting with environments. For instance, the authors in \cite{Wang2025a} examined communication security in ISAC-enabled low-altitude wireless networks and introduce new performance indicators, including data freshness, together with a resource-optimization framework that jointly balances sensing and computing performance while enhancing the secrecy rate, thereby establishing a novel method for securing LAWNs.
\end{itemize}
\vspace{-12pt}
\subsection{Fundamental Limitations and Unsolved Problems}
\label{Subsec: Fundamental Limitations and Unsolved Problems}
\par Although traditional AI-enabled solutions have demonstrated effectiveness in addressing specific security threats in LAWNs, they encounter some fundamental limitations, which can be summarized as follows.

\begin{itemize}
    \item \textbf{\textit{Limited generalization across diverse scenarios}:} Most traditional AI models are developed for narrow and task-specific applications. When deployed in unseen environments that feature mobility patterns or adversarial behaviors, their performance often deteriorates. This significantly limits their ability to adapt to the heterogeneous and uncertain nature of LAWN operations.

    \item \textbf{\textit{Insufficient use of multi-modal sensing information}:} Conventional approaches typically rely on single-modal inputs and treat communication tasks in isolation. Thus, they fail to incorporate other valuable sources of information, such as environmental context and mission-level semantics. The lack of multi-modal integration limits a comprehensive understanding of the surrounding environment under partial observations.

    \item \textbf{\textit{Reactive behavior instead of proactive reasoning}:} Existing methods mainly focus on detecting and responding to attacks after they occur. For example, the authors in \cite{Wang2025b} constructed a Stackelberg game model for attacks on integrated sensing and communication networks and design dynamic resource-adjustment strategies to maintain system stability under adversarial conditions, thereby supporting secure and stable operation. However, they cannot often anticipate evolving threats, perform reasoning under uncertainty or generate proactive defense strategies. This limitation reduces their effectiveness against stealthy, adaptive, or coordinated attacks in low-altitude airspace.
\end{itemize}

\vspace{-13pt}
%
%
\section{LAM-enabled Secure Communications in LAWNs}
\label{Sec: LAM-enabled Secure Communications in LAWNs}

\begin{figure*}
    \centering
    \includegraphics[width=\linewidth]{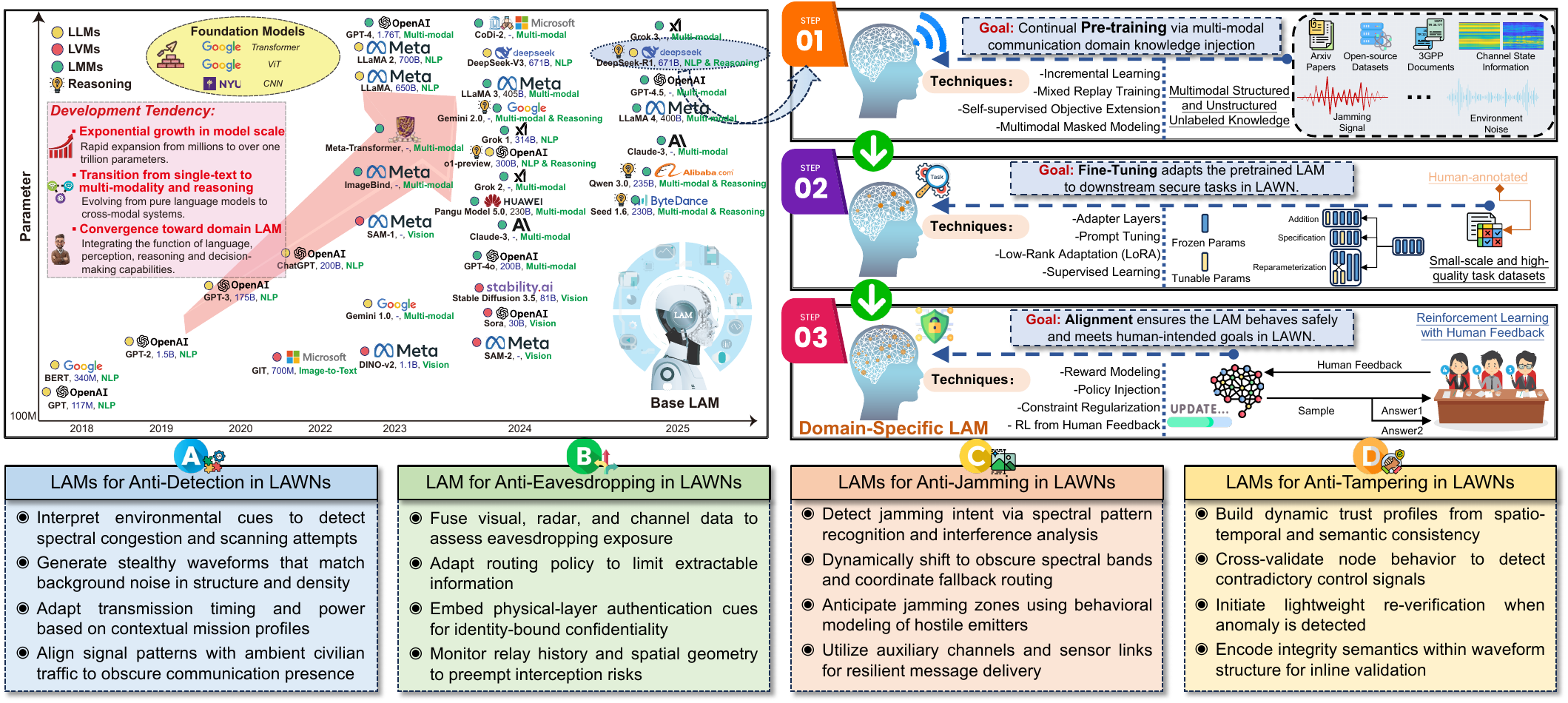}
    \caption{{\color{color_revise}{Overview and domain adaptation of LAMs, as well as their use in enabling secure communications in LAWNs.}}}
    \label{Fig: Mag_Fig_2}
    \vspace{-14pt}
\end{figure*}
\subsection{Overview of LAM}
\label{Subsec: Overview of LAM}

\subsubsection{Taxonomy of LAMs}
\label{Subsubsec: Taxonomy of LAMs}

\par Recent advances in the field of AI have driven a clear shift from narrow and task-specific models toward large-scale and general-purpose LAMs. These models are increasingly exhibiting cross-modal perception, contextual reasoning and generative capabilities, thereby enabling them to tackle a wide range of complex problems across various domains. {\color{color_revise}{Accordingly, LAMs have become powerful bases for generation, reasoning, and decision-making.}} Based on the types of input modalities, LAMs can be categorized as follows.

\begin{itemize}
    \item \textbf{\textit{LLMs:}} LLMs are foundation models trained on large-scale text corpora to understand, generate and interact with natural language. Thus, they excel in instruction following, summarization and knowledge retrieval. In principle, most LLMs are built upon the transformer architecture, which employs self-attention mechanisms to model contextual relationships between tokens across long sequences \cite{Vaswani2023}. This structure enables efficient parallel processing and the capture of global semantics, thereby supporting language-grounded reasoning, planning, and control when integrated with other modalities. Building upon these capabilities, a number of powerful LLMs have been developed in recent years. Among them, representative models include \href{https://github.com/openai/gpt-3}{GPT-3} and \href{https://github.com/meta-llama/llama}{LLaMA 2}.

    \item \textbf{\textit{Large vision models (LVMs):}} LVMs are foundation models designed to process high-dimensional visual data, including images and video streams. They support tasks such as object detection, semantic segmentation, motion tracking and scene understanding. Architecturally, most LVMs are built upon convolutional neural networks (CNNs) and visual transformers (ViTs), which enable the extraction of hierarchical features that capture both fine-grained structures and global spatial context \cite{Han2023}. These models often rely on large-scale annotated datasets or self-supervised pretraining to learn transferable visual representations. Thus, LVMs exhibit strong perception and generation capabilities in image analysis and synthesis. As typical models of LVMs, \href{https://github.com/CompVis/stable-diffusion}{Stable Diffusion} and \href{http://openai.com/sora/}{Sora} are highly capable in visual understanding.

    \item \textbf{\textit{Large multi-modal models (LMMs):}} LMMs are foundation models developed to jointly process and understand inputs from multiple modalities, such as text, images, audio and sensor data. They are typically constructed using modality-specific encoders to extract structured features, projection modules to align these features into a shared representation space and a central reasoning unit based on an LLM semantic understanding and task execution. This architecture enables LMMs to perform effective cross-modal fusion and support tasks including visual question answering, multi-modal instruction following and context-aware decision-making. Based on these multi-modal capabilities, several powerful LMMs have emerged in recent years. Representative models include \href{https://openai.com/index/hello-gpt-4o/}{GPT-4o} and \href{https://codi-2.github.io/}{CoDi-2}. Moreover, more advanced models, such as \href{https://ai.meta.com/blog/llama-4-multimodal-intelligence/}{LLaMA 4}, integrate the reasoning mechanisms to support unified perception and cognition across modalities.

\end{itemize}


\subsubsection{From Base LAMs to Domain-specific LAMs}
\label{Subsubsec: From Base LAMs to Domain-specific LAMs}

\par To bridge the gap between general-purpose intelligence and domain-specific applications, LAMs typically undergo a structured learning pipeline that incrementally aligns their foundational capabilities with specialized task demands \cite{Zhu2025}. As illustrated in Fig.~\ref{Fig: Mag_Fig_2}, domain-specific LAMs are developed through a progressive adaptation process, which can be exemplified by secure communications in LAWNs as follows.

\begin{itemize}
    \item \textbf{\textit{Pre-training}:} The pre-training stage equips the model with foundational generative and perceptual capabilities in LAWNs by leveraging large-scale and diverse datasets. Representative inputs at this stage include open-source datasets, 3GPP protocol documents, and frequently encountered jamming patterns, among others. These inputs are intended to reflect the dynamic characteristics of real-world LAWN environments and can be extended to other relevant modalities as required. In this phase, the model incorporates techniques such as incremental learning, mixed replay training, self-supervised objective extension and multimodal masked modeling. These approaches enable the LAM to learn semantic structures, capture cross-modal correlations and build foundational capabilities for downstream adaptation in LAWNs.
    
    \item \textbf{\textit{Fine-tuning}:} The fine-tuning stage adapts the pre-trained model to domain-specific tasks and operational characteristics of secure communications in LAWNs. Typical inputs include small-scale and high-quality annotated datasets consisting of simulation data under various channel conditions and expert-labeled security scenarios. These inputs focus on task-level distinctions and environmental variability that are not fully captured during the pre-training. In this phase, techniques such as prompt tuning, adapter modules, low-rank adaptation and supervised fine-tuning can be employed. These methods adjust the internal representations of the pre-trained model to emphasize task-relevant knowledge while preserving generalization from pre-training. As a result, the LAM gains the capability to perform specialized tasks such as jamming avoidance, eavesdropping detection and covert waveform generation under real-time and dynamic conditions of LAWN environments.

    \item \textbf{\textit{Alignment}:} The alignment stage ensures that the behavior of the fine-tuned model remains consistent with mission objectives, operational safety constraints and human-aligned preferences. Inputs at this stage include domain-specific reward signals, mission-level goals and feedback from real-world interactions. Unlike fine-tuning, alignment emphasizes value alignment, safety and interpretability in complex environments. Core techniques include reward modeling, constraint-aware policy optimization, RL with human feedback and rule-based behavior regularization. These approaches guide the model toward making context-aware decisions that balance competing objectives such as secure communication efficiency and communication covertness of LAPs.
\end{itemize}

\par A comparison between LAMs and traditional AI models is summarized in Table I. {\color{color_revise}{Moreover, model optimization strategies such as pruning, early exits, and knowledge distillation can significantly reduce the computational burden and energy consumption, making LAMs more efficient for deployment in LAWNs.}}

\begin{table*}[ht]
\centering
\arrayrulecolor{white}
\caption{{\color{color_revise}{Comparison of LAMs with Traditional AI Models}}}
\vspace{-5pt}
\label{tab:AI_comparison_vertical}
\renewcommand{\arraystretch}{1.1}

\begin{tabular}{
|>{\columncolor{colDim1}}>{\centering\arraybackslash}m{4cm}
|>{\columncolor{colTrad1}}>{\raggedright\arraybackslash}m{6.35cm}
|>{\columncolor{colLAM1}}>{\raggedright\arraybackslash}m{6.35cm}|}
\hline
\cellcolor{colDim1} \textbf{\textit{Aspects}} &
\multicolumn{1}{>{\centering\arraybackslash}m{6.35cm}|}{\cellcolor{colTrad1}\textit{\textbf{Traditional AI Models}}} &
\multicolumn{1}{>{\centering\arraybackslash}m{6.35cm}|}{\cellcolor{colLAM1}\textit{\textbf{LAMs}}} \\
\hline
\textbf{Model Structure} &
Focus on narrow and task-specific objectives with simple architectures &
Support diverse downstream tasks via unified and general-purpose architectures \\
\hline
\textbf{Learning Paradigm} &
Learn from small or static datasets via supervised, unsupervised or RL &
Learn progressively through large-scale pre-training, domain fine-tuning and alignment \\
\hline
\textbf{Modality Handling} &
Process single-modality input with handcrafted fusion strategies &
Fuse multi-modal inputs through end-to-end semantic learning mechanisms \\
\hline
\textbf{Generalization and Adaptability} &
Struggle to adapt in unseen conditions or environments &
Generalize across diverse scenarios with strong robustness and transferability \\
\hline
\textbf{Reasoning Capability} &
React to patterns with limited or no reasoning ability &
Perform multi-step reasoning with contextual abstraction and task planning \\
\hline
\hline
\end{tabular}
\vspace{-14pt}
\end{table*}

\vspace{-12pt}
\subsection{Roles of LAMs for Secure Communications in LAWNs}
\label{Subsec: Roles of LAMs for Secure Communications in LAWNs}

\par After undergoing domain adaptation, LAMs can contribute to various aspects of secure communications in LAWNs.

\begin{itemize}
    \item \textbf{\textit{LAMs for Anti-detection}:}  
    LAMs excel in anti-detection by leveraging their generative modeling capabilities and context-aware inference to produce stealthy waveforms that seamlessly integrate with ambient electromagnetic noise. Traditional AI models, which are often trained for narrow tasks using static datasets, cannot reason over mission context or environmental dynamics, resulting in communication behaviors that are more predictable and thus vulnerable to detection. In contrast, LAMs can interpret environmental cues, such as spectral congestion, timing anomalies or adversarial scanning patterns, and generate transmission signals that adapt in both structure and spectral appearance. During low-altitude urban flights, an LLM-enabled LAP can identify elevated background noise from residential wireless devices. It then synthesizes communication signals with matching spectral density and intermittent burst structure, thereby allowing transmissions to blend with transient emissions from some other traffic to reduce the detection.

    \item \textbf{\textit{LAMs for Anti-eavesdropping}:}  
    LAMs enhance anti-eavesdropping capability by using multi-modal sensing and adaptive control of signal structures and routing. By reasoning over spatial geometry, relay behaviors, and channel observability, LAM-enabled LAP nodes can detect high-risk interception scenarios and proactively adjust routing paths or transmission formats to reduce information leakage. {\color{color_revise}{Moreover, semantic steganography techniques proposed in \cite{Bai2024} allow LLMs to hide information at the semantic level within generated content, making intercepted messages difficult to interpret even when captured.}} Combining semantic-level hiding with physical-layer adaptation forms a layered defense mechanism, enabling LAM-enabled LAPs to effectively suppress eavesdropping threats while maintaining reliable communication in LAWNs.

    \item \textbf{\textit{LAMs for Anti-jamming}:}  
    LAMs provide robust jamming resistance by leveraging their decision-making and multi-modal fusion capabilities to orchestrate agile and resilient communication strategies. Unlike conventional systems that follow static avoidance or fallback protocols, LAMs continuously assess jamming intensity, spectral patterns, and communication urgency to determine the optimal response. For example, in a dynamic low-altitude scenario where a mobile LAP cluster encounters burst jamming across key control frequencies, an LLM-enabled LAP can infer the jamming intent and rapidly shift communication to a more obscure spectral band while coordinating fallback messaging through indirect paths, such as auxiliary sensor links or side channels. Additionally, by modeling the behavior of nearby devices, an LLM-enabled LAP can preemptively predict potential interference zones and adjust spectrum use before disruption occurs. This anticipatory and coordinated approach enables LAWNs to maintain control and information flow under hostile spectrum conditions.

    \item \textbf{\textit{LAMs for Anti-tampering}:}  
    {\color{color_revise}{Beyond traditional verification mechanisms, LAMs strengthen anti-tampering in LAWNs through adaptive trust modeling and federated semantic verification. Similar to FedLLMGuard proposed in \cite{Rezaei2025}, LAMs move beyond fixed keys or static signatures by constructing dynamic trust profiles based on spatiotemporal consistency, semantic coherence, traffic context, and historical node behavior. When a ground or aerial relay node issues anomalous or semantically conflicting control messages, the LAM cross-checks these updates with observations from neighboring nodes to detect signs of tampering or coordinated manipulation. Moreover, LAMs support decentralized and privacy-preserving trust validation. Instead of discarding abnormal messages outright, the LAM may initiate a lightweight and role-aware re-verification process to reduce false positives in dynamic LAWN environments.}}
\end{itemize}


\vspace{-12pt}
\subsection{Lessons Learned}
\label{Subsec: Lessons Learned}
\par {\color{color_revise}{Through the exploration of LAMs within LAWNs, the superiority of LAMs in complex low-altitude environments stems from their advanced generative and contextual reasoning abilities. These strengths enable LAMs to construct high-level abstract representations, interpret mission objectives, infer adversarial intent, and understand operational constraints in a manner unattainable for conventional AI models.}}
\vspace{-12pt}
%
%
\section{Case Study}
\label{Sec: Case Study}
\begin{figure*}
    \centering
    \includegraphics[width=\linewidth]{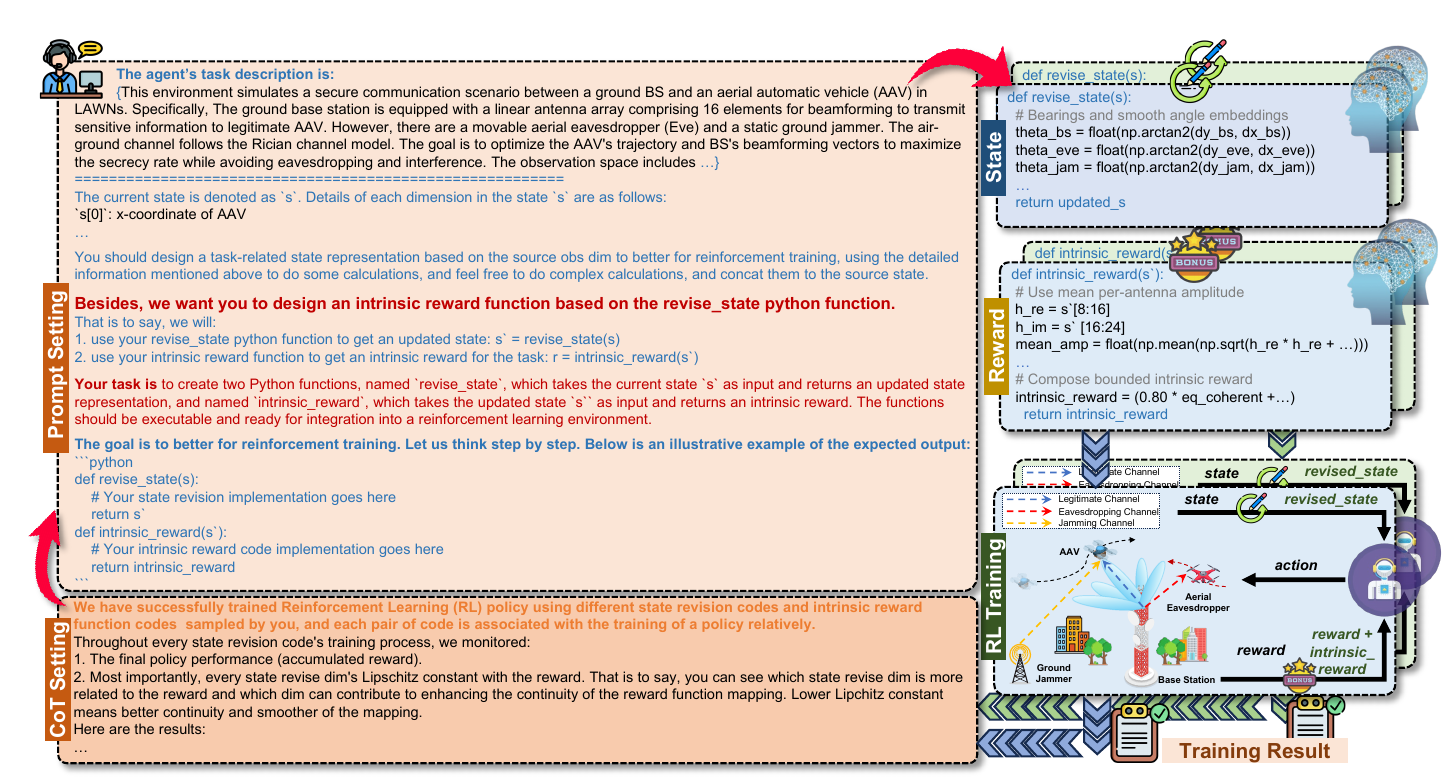}
    \caption{{\color{color_revise}{CoT-LLM-enhanced state representation and reward function design for RL toward secure communications in LAWNs.}}}
    \label{Fig: Mag_Fig_3}
    \vspace{-14pt}
\end{figure*}

\begin{figure}
    \centering    \includegraphics[width=0.95\linewidth]{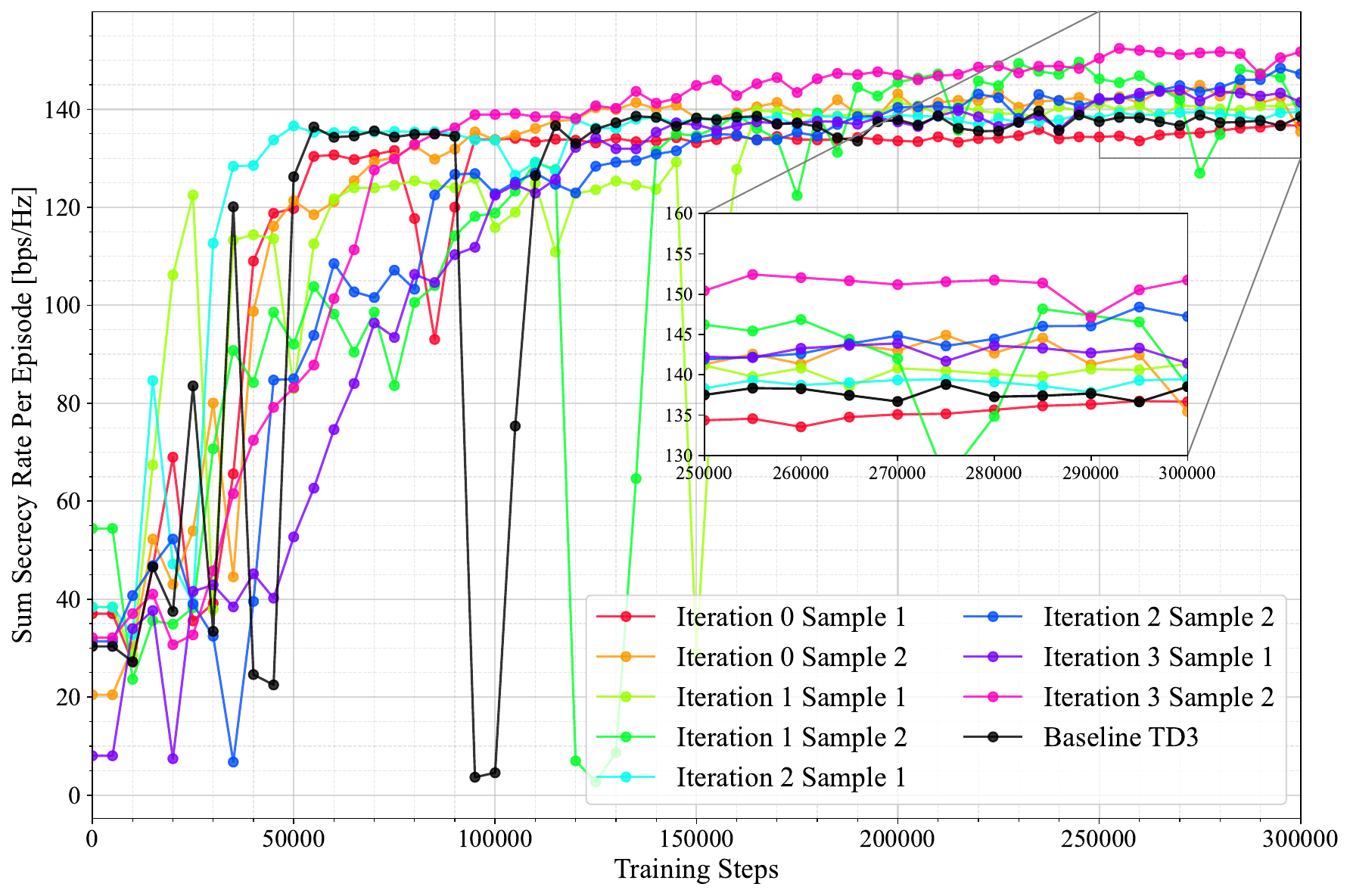}
    \caption{Performance comparison for the proposed framework and baseline.}
    \label{Fig: Result}
    \vspace{-14pt}
\end{figure}

{\color{color_revise}{\par This section presents a CoT-LLM-enhanced RL framework to improve decision-making for secure communications in LAWNs.
\vspace{-10pt}
\subsection{CoT-LLM-Enhanced Optimization Framework Design}
\label{Subsec: LLM-Enhanced Optimization Framework Design}

\par The proposed framework brings into practice essential capabilities of LAMs, especially their strengths in generative modeling and context-aware reasoning \cite{Peng2024}. As a representative subclass of LAMs, we instantiate these advantages through a CoT-enabled LLM tailored to enhance RL for secure communications in LAWNs. As shown in Fig~\ref{Fig: Mag_Fig_3}, the process of integrating the CoT-LLM into the RL is outlined as follows.

\begin{itemize}
    \item \textbf{\textit{Step 1: LLM-generated candidates.}}  
    At the beginning of each iteration, the LLM receives a symbolic description of the manually defined observation space together with the objective. Based on this information, it produces multiple candidate state representations and corresponding intrinsic reward function designs.

    \item \textbf{\textit{Step 2: Parallel RL evaluation.}}  
    Each state representation and reward function candidate is then integrated into a separate RL learner and trained. By observing the learning trajectories produced under these different designs, the framework derives a performance profile for each design based on its reward evolution and the smoothness of its induced feature dynamics, where the smoothness is reflected by the empirical Lipschitz constant and smaller values between the feature and the reward indicate more stable and reliable learning behavior.

    \item \textbf{\textit{Step 3: Feedback extraction.}}  
    After the training evaluation, structured feedback is extracted from the learning processes of all candidates. This feedback captures reward evolution and the consistency of enhanced features.

    \item \textbf{\textit{Step 4: CoT-based refinement.}}  
    The collected feedback is converted into concise textual instructions and fed back to the LLM. Through CoT reasoning, the LLM interprets this feedback and refines its earlier state representation and intrinsic designs by removing unstable components, strengthening informative features, and reorganizing reward structures to better align with objectives.

    \item \textbf{\textit{Step 5: Closed-loop iteration.}}  
    The refined state and reward designs are used to generate a new set of candidates for the next iteration, thus forming a closed-loop process. Over successive rounds, the CoT-enabled LLM produces increasingly stable and semantically meaningful representations, enabling improved performance in LAWNs.
\end{itemize}

\vspace{-15pt}
\subsection{Scenario Description}
\label{Subsec: Scenario Description}

\par We consider a typical secure communication scenario in LAWNs, where a ground base station (BS) equipped with a uniform linear array (ULA) consisting of 16 antenna elements transmits confidential information to an aerial autonomous vehicle (AAV) within a task area. A movable aerial eavesdropper attempts to intercept the transmission, while a ground jammer operating with randomly varying power emits interference. The air-to-ground channels follow a Rician channel model, and the objective is to jointly optimize the trajectory of the AAV and the beamforming vectors of the BS to maximize the sum secrecy rate under both eavesdropping and jamming threats. Additional information on the simulation procedures is available on our project website: https://chuangzhang1999.github.io/LAM4SCLAWN.github.io/.
\vspace{-15pt}
\subsection{Numerical Results}
\label{Subsec: Numerical Results}

\par For the proposed CoT-LLM-enhanced optimization framework, we adopt twin delayed deep deterministic policy gradient (TD3) as the base RL algorithm and GPT-5.1 as the underlying LLM to support semantic state construction and intrinsic reward shaping. Fig.~\ref{Fig: Result} shows that the proposed framework progressively improves convergence stability and secrecy rate across iterations. Iteration~0 behaves similarly to standard TD3, with clear early-stage fluctuations. Iteration~1 improves mid-training performance but still exhibits noticeable instability. From Iteration~2 onward, the learning process becomes more robust, and by Iteration~3 the curves become significantly smoother with consistently superior secrecy performance. These results indicate that the CoT-driven closed-loop feedback effectively guides the LLM to generate increasingly stable and meaningful state and reward designs.}}

%
%
\section{Future Directions}
\label{Sec: Future Directions}

\subsection{Construction of Domain-Specific and Multi-modal Datasets}
\label{Subsec: Construction of Domain-Specific and Multimodal Datasets}

\par The performance of LAMs in secure communications is closely tied to the quality and diversity of their training data. However, existing datasets often fall short in capturing the multi-modal and dynamic characteristics of realistic LAWN scenarios. Future work should focus on constructing large-scale, domain-specific and multi-modal datasets that incorporate wireless signal traces, mobility patterns, jamming behaviors and environment-aware parameters.
{\color{color_revise}{
\subsection{Efficient and Privacy LAM for Distributed Deployment}
\label{Subsec: Efficient LAM for Distributed Deployment}

\par Despite their powerful capabilities, LAMs are often too large and resource-intensive for direct deployment on aerial platforms with limited computational capacity. Moreover, privacy preservation is also a critical concern in distributed environments. Future work could focus on optimizing model architectures through dynamic pruning and quantization and edge computing techniques, enabling more efficient execution. Furthermore, federated learning can be incorporated into LAMs to ensure that sensitive data is never shared directly.}}

\subsection{Trustworthy Reasoning in Adversarial Environments}
\label{Subsec: Trustworthy Reasoning in Adversarial Environments}

\par As LAMs are integrated into mission-critical security functions, ensuring predictable and safe behavior under adversarial and uncertain conditions becomes essential. Future research should prioritize the development of safety-aligned reasoning mechanisms, such as RL with safety constraints and interpretable decision processes.

%
%
\section{Conclusion}
\label{Sec: Conclusion and Future Directions}

\par In this paper, we have explored the potential of LAMs to enhance secure communications in LAWNs. Specifically, we have identified the unique security challenges faced by LAWNs due to its reliance on LAPs and unlicensed spectrum, and discussed the limitations of traditional AI methods in addressing these challenges. By integrating LAMs into LAWNs, we have examined their capabilities across diverse roles within the network, thereby enabling the design of adaptive, proactive, and robust security mechanisms suited to each context. {\color{color_revise}{Moreover, we have proposed an LAM-based optimization framework that leverages LLMs with CoT reasoning to iteratively refine RL state representations and reward design toward secure communications in LAWNs, which has been validated through a case study.}} Furthermore, several directions have been marked to provide guidance.

\ifCLASSOPTIONcaptionsoff
\newpage
\fi

\bibliographystyle{IEEEtran}
\bibliography{references.bib}

\begin{thebibliography}{10}
\providecommand{\url}[1]{#1}
\csname url@samestyle\endcsname
\providecommand{\newblock}{\relax}
\providecommand{\bibinfo}[2]{#2}
\providecommand{\BIBentrySTDinterwordspacing}{\spaceskip=0pt\relax}
\providecommand{\BIBentryALTinterwordstretchfactor}{4}
\providecommand{\BIBentryALTinterwordspacing}{\spaceskip=\fontdimen2\font plus
\BIBentryALTinterwordstretchfactor\fontdimen3\font minus \fontdimen4\font\relax}
\providecommand{\BIBforeignlanguage}[2]{{%
\expandafter\ifx\csname l@#1\endcsname\relax
\typeout{** WARNING: IEEEtran.bst: No hyphenation pattern has been}%
\typeout{** loaded for the language `#1'. Using the pattern for}%
\typeout{** the default language instead.}%
\else
\language=\csname l@#1\endcsname
\fi
#2}}
\providecommand{\BIBdecl}{\relax}
\BIBdecl

\bibitem{Wang2025}
Y.~Wang, G.~Sun, Z.~Sun, J.~Wang, J.~Li, C.~Zhao, J.~Wu, S.~Liang, M.~Yin, P.~Wang, D.~Niyato, S.~Sun, and D.~In~Kim, ``Toward realization of low-altitude economy networks: Core architecture, integrated technologies, and future directions,'' \emph{{IEEE} Trans. Cognit. Commun. Networking}, vol.~11, no.~5, pp. 2788--2820, 2025.

\bibitem{Cai2025}
L.~Cai, J.~Wang, R.~Zhang, Y.~Zhang, T.~Jiang, D.~Niyato, X.~Wang, A.~Jamalipour, and X.~Shen, ``Secure physical layer communications for low-altitude economy networking: A survey,'' \emph{{IEEE} Commun. Surv. Tutorials}, 2025, {E}arly Access, doi: {10.1109/COMST.2025.3634768}.

\bibitem{Jiang2025}
F.~Jiang, C.~Pan, L.~Dong, K.~Wang, O.~A. Dobre, and M.~Debbah, ``From large {AI} models to agentic {AI}: A tutorial on future intelligent communications,'' \emph{arXiv preprint arXiv:2505.22311}, May 2025, doi: {10.48550/arXiv.2505.22311}.

\bibitem{Yang2025}
T.~Yang, P.~Zhang, M.~Zheng, Y.~Shi, L.~Jing, J.~Huang, and N.~Li, ``{WirelessGPT}: A generative pre-trained multi-task learning framework for wireless communication,'' \emph{arXiv preprint arXiv:2502.06877}, Feb. 2025, doi: {10.48550/arXiv.2502.06877}.

\bibitem{Zhang2025}
R.~Zhang, W.~Wu, X.~Chen, Z.~Gao, and Y.~Cai, ``Terahertz integrated sensing and communication-empowered {UAVs} in {6G}: A transceiver design perspective,'' \emph{{IEEE} Veh. Technol. Mag.}, pp. 2--11, 2025, {E}arly Access, doi:{10.1109/MVT.2025.3531088}.

\bibitem{Vo2025}
V.~N. Vo, N.~Q. Long, V.~Dang, T.~D. Ho, H.~Tran, S.~Chatzinotas, D.~H. Tran, S.~Sanguanpong, and C.~So{-}In, ``Deep learning-driven throughput maximization in covert communication for {UAV-RIS} cognitive systems,'' \emph{{IEEE} Open J. Commun. Soc.}, vol.~6, pp. 4140--4155, Apr. 2025.

\bibitem{Heo2024}
K.~Heo, W.~Lee, and K.~Lee, ``{UAV}-assisted wireless-powered secure communications: Integration of optimization and deep learning,'' \emph{{IEEE} Trans. Wirel. Commun.}, vol.~23, no.~9, pp. 10\,530--10\,545, Sep. 2024.

\bibitem{Wang2025a}
J.~Wang, C.~Zhao, J.~He, G.~Sun, W.~Yuan, D.~Niyato, L.~Zhu, and T.~Xiang, ``Security-aware joint sensing, communication, and computing optimization in low altitude wireless networks,'' \emph{arXiv preprint arXiv:2511.01451}, Nov. 2025, doi: {10.48550/arXiv.2511.01451 }.

\bibitem{Wang2025b}
J.~Wang, C.~Zhao, D.~Niyato, G.~Sun, W.~Yuan, A.~Jamalipour, and T.~Xiang, ``Stackelberg game-driven defense for {ISAC} against channel attacks in low-altitude networks,'' \emph{arXiv preprint arXiv:2511.06359}, Nov. 2025, doi: {10.48550/arXiv.2511.06359}.

\bibitem{Vaswani2023}
A.~Vaswani, N.~Shazeer, N.~Parmar, J.~Uszkoreit, L.~Jones, A.~N. Gomez, L.~Kaiser, and I.~Polosukhin, ``Attention is all you need,'' \emph{arXiv preprint arXiv:1706.03762}, Aug. 2023, doi: {10.48550/arXiv.1706.03762}.

\bibitem{Han2023}
K.~Han, Y.~Wang, H.~Chen, X.~Chen, J.~Guo, Z.~Liu, Y.~Tang, A.~Xiao, C.~Xu, Y.~Xu, Z.~Yang, Y.~Zhang, and D.~Tao, ``A survey on vision transformer,'' \emph{{IEEE} Trans. Pattern Anal. Mach. Intell.}, vol.~45, no.~1, pp. 87--110, Jan. 2023.

\bibitem{Zhu2025}
F.~Zhu, X.~Wang, X.~Li, M.~Zhang, Y.~Chen, C.~Huang, Z.~Yang, X.~Chen, Z.~Zhang, R.~Jin, Y.~Huang, W.~Feng, T.~Yang, B.~Bai, F.~Gao, K.~Yang, Y.~Liu, S.~Muhaidat, C.~Yuen, K.~Huang, K.-K. Wong, D.~Niyato, and M.~Debbah, ``Wireless large {AI} model: Shaping the {AI}-native future of {6G} and beyond,'' \emph{arXiv preprint arXiv:2504.14653}, Apr. 2025, doi: {10.48550/arXiv.2504.14653}.

\bibitem{Bai2024}
M.~Bai, J.~Yang, K.~Pang, Y.~Huang, and Y.~Gao, ``Semantic steganography: A framework for robust and high-capacity information hiding using large language models,'' \emph{arXiv preprint arXiv:2412.11043}, Dec. 2024, doi: {10.48550/arXiv.2412.11043}.

\bibitem{Rezaei2025}
H.~Rezaei, R.~Taheri, and M.~Shojafar, ``Fedllmguard: A federated large language model for anomaly detection in 5g networks,'' \emph{Comput. Networks}, vol. 269, p. 111473, 2025.

\bibitem{Peng2024}
I.~S. Andi~Peng, T.~R.~S. Belinda Z.~Li, T.~L. Griffiths, J.~Andreas, and J.~A. Shah, ``Learning with language-guided state abstractions,'' in \emph{Proc. 12th Int. Conf. Learn. Representations (ICLR)}, Vienna, Austria, May 7-11, 2024, pp. 1--34.

\end{thebibliography}

\section*{Biographies}

\noindent 
\textsc{Chuang Zhang} (\text{chuangzhang1999@gmail.com}) received the B.S. degree in computer science and technology from Jilin University, Changchun, China, in 2021, where he is currently pursuing the Ph.D. degree with the College of Computer Science and Technology. He was also a Visiting Student with the Singapore University of Technology and Design, Singapore. His current research interests include UAV communications, secure communications, distributed beamforming and multi-objective optimization.
\vspace{1em}

\noindent 
\textsc{Geng Sun} (\text{sungeng@jlu.edu.cn}) received the B.S. degree in communication engineering from Dalian Polytechnic University, and the Ph.D. degree in computer science and technology from Jilin University, in 2011 and 2018, respectively. He was a Visiting Researcher with the School of Electrical and Computer Engineering, Georgia Institute of Technology, USA. He is a Professor in the College of Computer Science and Technology at Jilin University. Currently, he is working as a visiting scholar at the College of Computing and Data Science, Nanyang Technological University, Singapore. He has published over 100 high-quality papers, including IEEE TMC, IEEE JSAC, IEEE/ACM ToN, IEEE TWC, IEEE TCOM, IEEE TAP, IEEE IoT-J, IEEE TIM, IEEE INFOCOM, IEEE GLOBECOM, and IEEE ICC. He serves as the Associate Editors of IEEE Communications Surveys \& Tutorials, IEEE Transactions on Communications, IEEE Transactions on Vehicular Technology, IEEE Transactions on Network Science and Engineering, IEEE Transactions on Network and Service Management and IEEE Networking Letters. He serves as the Lead Guest Editor of Special Issues for IEEE Transactions on Network Science and Engineering, IEEE Internet of Things Journal, IEEE Networking Letters. He also serves as the Guest Editor of Special Issues for IEEE Transactions on Services Computing, IEEE Communications Magazine, and IEEE Open Journal of the Communications Society. His research interests include Low-altitude Wireless Networks, UAV communications and Networking, Mobile Edge Computing (MEC), Intelligent Reflecting Surface (IRS), Generative AI and Agentic AI, and deep reinforcement learning.

\vspace{1em}

\noindent 
\textsc{Yijing Lin} (\text{yjlin@bupt.edu.cn}) received the Ph.D. degree from the State Key Laboratory of Networking and Switching Technology, Beijing University of Posts and Telecommunications (BUPT), Beijing, China, in 2024. She is currently a Post-Doctoral Researcher with the State Key Laboratory of Networking and Switching Technology, BUPT. She has published more than 20 papers, including WWW, IJCAI, IEEE Transactions on Services Computing, IEEE Transactions on Communications, IEEE Transactions on Network Science and Engineering, and IEEE Transactions on Vehicular Technology. Her publications include ESI highly cited papers and IEEE ComSoc Best Readings. Her current research interests include blockchain and data unlearning.

\vspace{1em}

\noindent 
\textsc{Weijie Yuan} (\text{yuanwj@sustech.edu.cn}) is currently an Assistant Professor with the Southern University of Science and Technology. His research interests include integrated sensing and communications (ISAC), orthogonal time frequency space (OTFS), and low-altitude wireless networks (LAWN). He was a recipient of the Best Editor from IEEE CommL; the Best Paper Award from IEEE ICC 2023, IEEE/CIC ICCC 2023, and IEEE GlobeCom 2024; and the 2025 IEEE Communications Society and Information Theory Society Joint Paper Award. He was the Track-Chair of IEEE ICC 2025 and IEEE VTC 2025-Spring. He served as an Organizer/the Chair for several workshops and special sessions in flagship IEEE and ACM conferences, including IEEE ICC, IEEE VTC, IEEE GlobeCom, IEEE/CIC ICCC, IEEE SPAWC, IEEE WCNC, IEEE ICASSP, and ACM MobiCom. He is the Founding Chair of the IEEE ComSoc Special Interest Group (SIG) on LAWN and the SIG on OTFS. He serves as an Editor for IEEE Transactions on Communications, IEEE Transactions on Wireless Communications, IEEE Transactions on Mobile Computing, IEEE Communications Magazine, IEEE Communications Standards Magazine, IEEE Transactions on Green Communications and Networking, IEEE Communications Letter, and IEEE Open Journal of Communications Society; and a Guest Editor for IEEE Transactions on Vehicular Technology, IEEE Transactions on Network Science and Engineering, and IEEE Internet of Things Journal. He was listed among the World’s Top 2\% Scientists by Stanford University for citation impact from 2021 to 2024 and among the Elsevier Highly-Cited Chinese Researchers.

\vspace{1em}

\noindent 
\textsc{Sinem Coleri} (\text{scoleri@ku.edu.tr}) received the B.S. degree in electrical and electronics engineering from Bilkent University, Ankara, Türkiye, in 2000, and the M.S. and Ph.D. degrees in electrical engineering and computer sciences from the University of California at Berkeley, Berkeley, CA, USA, in 2002 and 2005, respectively. She worked as a Research Scientist at the Wireless Sensor Networks Berkeley Laboratory under the sponsorship of Pirelli and Telecom Italia from 2006 to 2009. Since September 2009, she has been a Faculty Member with the Department of Electrical and Electronics Engineering, Koç University, Istanbul, Türkiye, where she is currently a Professor. She is also the Founding Director of the Wireless Networks Laboratory (WNL) and the Director of the Ford Otosan Automotive Technologies Laboratory. Her research interests include 6G wireless communications and networking, machine learning for wireless networks, machine-to-machine communications, wireless networked control systems, and vehicular networks. She has received numerous awards and recognitions, including the Scientific and Technological Research Council of Türkiye (TÜBİTAK) Science Award in 2024, the N2Women: Stars in Computer Networking and Communications in 2022, the TÜBİTAK Incentive, the IEEE Vehicular Technology Society Neal Shepherd Memorial Best Propagation Paper Award in 2020, the Outstanding Achievement Award by the Higher Education Council in 2018, and Turkish Academy of Sciences Distinguished Young Scientist (TUBA-GEBIP) Award in 2015. She serves as the Editor-in-Chief for IEEE Open Journal of the Communications Society and is an IEEE ComSoc Distinguished Lecturer.

\vspace{1em}

\noindent 
\textsc{Dusit Niyato} (\text{dniyato@ntu.edu.sg}) received the B.Eng. degree from the King Mongkuts Institute of Technology Ladkrabang (KMITL), Thailand, in 1999, and the Ph.D. degree in electrical and computer engineering from the University of Manitoba, Canada, in 2008. He is currently a Professor with the School of Computer Science and Engineering, Nanyang Technological University, Singapore. His research interests include the Internet of Things (IoT), machine learning, and incentive mechanism design.

\vfill

\end{document}